\documentclass[runningheads]{svmult}
\usepackage{makeidx}   % allows index generation
\usepackage{graphicx}  % standard LaTeX graphics tool
\usepackage{subeqnar}  % subnumbers individual equations
\usepackage{multicol}  % used for the two-column index
\usepackage{physprbb}  % modified textarea for proceedings,
\makeindex             % used for the subject index
\usepackage{epstopdf}
\usepackage{amsfonts}   % useful for coding complex math
\usepackage{amsbsy}
\usepackage{amsmath}
\usepackage{revsymb}	
\usepackage{natbib}
\bibpunct{(}{)}{;}{a}{}{,}

\def\note #1]{{\bf #1]}}
\def\fig{.}
\def\dd{{\rm d}}
\def\muHz{\,\mu{\rm Hz}}
\def\bolddelta{\delta\kern-0.45em\delta\kern-0.45em\delta}
\def\boldr{\mbox{\boldmath$r$}}
\def\bolddelr{\bolddelta \boldr}
\def\Msun{\,{\rm M_\odot}}
\def\Rsun{\,{\rm R_\odot}}

\begin{document}
\title*{Asteroseismology with solar-like oscillations}
\toctitle{Asteroseismology}
% allows explicit linebreak for the table of content
%
\titlerunning{Asteroseismology}
% allows abbreviation of title, if the full title is too long
% to fit in the running head
%
\author{J{\o}rgen Christensen-Dalsgaard}

\institute{Stellar Astrophysics Centre \\
Department of Physics and Astronomy, Aarhus University \\
Ny Munkegade 120, 8000 Aarhus C, Denmark \\
E-mail: \texttt{jcd@phys.au.dk}}

%\date{September 30, 2014}

\maketitle              % typesets the title of the contribution

\begin{abstract}
Almost 100 years ago Sir Arthur Eddington noted that the interiors of stars 
were inaccessible to observations.
The advent of helio- and asteroseismology has completely changed 
this assessment.
Helioseismology has provided very detailed information about 
the interior structure and dynamics of the Sun,
highlighting remaining issues in our understanding of the solar interior.
In the last decade extensive observations of stellar oscillations,
in particular from space photometry, have provided very detailed information
about the global and internal properties of stars. 
Here I provide an overview of these developments,
including the remarkable insight that has been obtained on the properties 
of evolved stars.
\end{abstract}

\section{Introduction}

In his seminal book {\it The internal constitution of the stars}
\citet{Edding1926} pondered `What appliance can pierce through the outer
layers of a star and test the conditions within?'.
He answered the question through his theoretical investigations of stellar
interiors which, despite the limited physical knowledge at the time, were
remarkably successful in uncovering the basic principles underlying
stellar internal structure.
Modelling stellar structure and evolution has undoubtedly become much more
sophisticated and, presumably, realistic since Eddington's time, but
the basic question remains: which {\it observations} can pierce through 
the outer layers of a star and test the conditions within?
The answer is now well-established: the study of stellar interiors through
the observation of oscillations on their surface, in other words,
{\it asteroseismology}.

\begin{figure}[!]
\begin{center}
	\includegraphics[width=8.7cm]{\fig/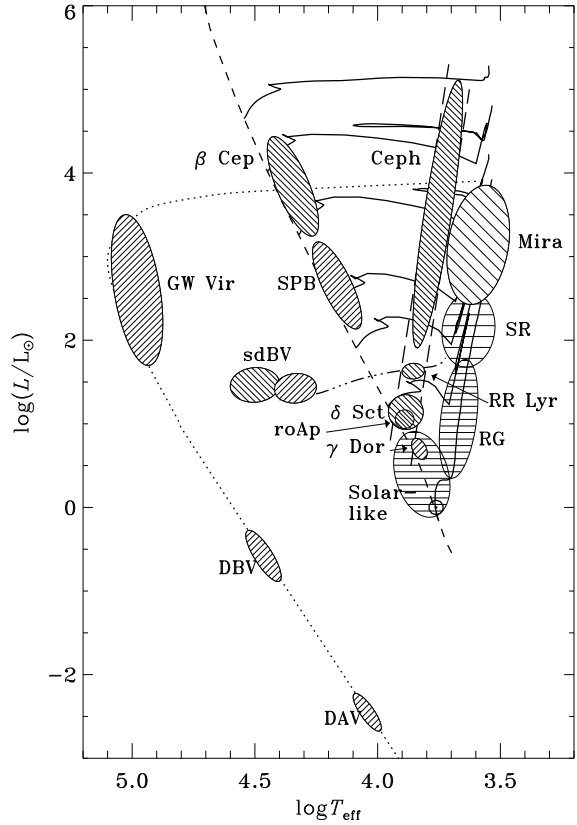}
\caption{Hertzsprung-Russell diagram showing the location of various groups
of pulsating stars.
The dashed line shows the zero-age main sequence and the solid lines show
selected evolutionary tracks. 
The dotted line schematically indicates the white-dwarf cooling track.
Here I focus on solar-like pulsators, indicated by horizontal hatching and
situated to the right of the Cepheid instability strip (marked `Ceph').
For further details, see \citet{Aerts2010}.
}
\label{fig:pulshr}
\end{center}
\end{figure}

The detailed observational study of stellar interiors started 
with the development of {\it helioseismology},
from extensive observations of oscillations on
the solar surface \citep[for a review, see, e.g.,][see also Kosovichev,
this volume]{Christ2002}.
However, as indicated in Fig.\ \ref{fig:pulshr} oscillations are found in a broad
range of stars, providing opportunities for studies of stars in essentially
all phases of their evolution.
In this brief review I focus on stars showing oscillations similar to
those of the Sun which, as discussed below, are intrinsically stable
and excited stochastically by the near-surface convection.

Modes of solar-like oscillations are generally characterized by extremely small
amplitudes, in the solar case up to about 20\,cm\,s$^{-1}$ in radial
velocity and a few parts per million in intensity.
The difficult observations of such oscillations in distant stars 
had a modest beginning in the nineties
\citep[e.g.,][]{Brown1991, Kjelds1995, Bouchy2001},
but they have evolved dramatically in the last few years through space-based 
photometric observations from the CoRoT \citep{Baglin2013} and,
in particular, the NASA {\it Kepler} mission \citep{Boruck2010, Gillil2010}
launched in March 2009.
Thus we stand at the beginning of a new phase of strongly 
observationally constrained studies of stellar interiors.

\section{Basic properties of stellar oscillations}
\label{sec:properties}

Here I only give a brief overview of the main features of stellar oscillations.
For a detailed description, see \citet{Aerts2010}.

Stellar oscillations are characterized by the dominant restoring forces and
the mechanisms exciting the modes.
Another important characteristic is the geometrical structure of the mode.
In (nearly) spherically symmetric stars this is characterized by a 
spherical harmonic $Y_l^m(\theta, \phi)$ as a function of co-latitude $\theta$
and longitude $\phi$.
Here the degree $l$ measures the total number of nodal lines on the stellar
surface and the azimuthal order $m$, with $|m| \le l$,
gives the number of nodal lines crossing the equator.
Spherically symmetric modes, with $l=0$, are known as radial oscillations.
In addition, a mode is characterized by the radial order $n$ related, 
sometimes in a rather complex manner \citep{Takata2012},
to the number of nodes in the radial direction.
In distant stars, where only oscillations in light integrated over 
the stellar surface have so far been analysed,
cancellation suppresses the modes of higher degree, and
only modes up to typically $l = 2 - 3$ are observed.

The two dominant restoring forces are pressure, and buoyancy acting on 
density differences.
The pressure-driven modes, known as p modes,
are essentially standing sound waves.
These are the modes observed in the Sun and solar-like oscillations in
moderately evolved stars.
They tend to have relatively high frequency, in the solar case between
1000 and $5000 \muHz$, corresponding to periods between 17 and 3 minutes, and
high radial order.
Consequently, their properties are well characterized by their 
asymptotic behaviour, according to which, to leading order, the
cyclic frequency satisfies
\begin{equation}
\nu_{nl} \simeq \Delta \nu \left(n + {l \over 2} + \epsilon \right) \; ,
\label{eq:pasymp}
\end{equation}
where $\epsilon$ is a frequency-dependent phase largely determined by
the near-surface properties.
Here the {\it large frequency separation} is given by
\begin{equation}
\Delta \nu = \left(2 \int_0^r {\dd r \over c} \right)^{-1} \; ,
\label{eq:largesep}
\end{equation}
where $c$ is the adiabatic sound speed and the integral is over
distance $r$ to the centre, between the centre and the surface radius $R$.
It may be shown that $\Delta \nu$, and hence the frequencies, scales
as the square root of the stellar mean density,
$\Delta \nu \propto (M/R^3)^{1/2}$ \citep{Ulrich1986},
where $M$ is the mass of the star.

\begin{figure}[!]
\begin{center}
\includegraphics[width=8.7cm]{\fig/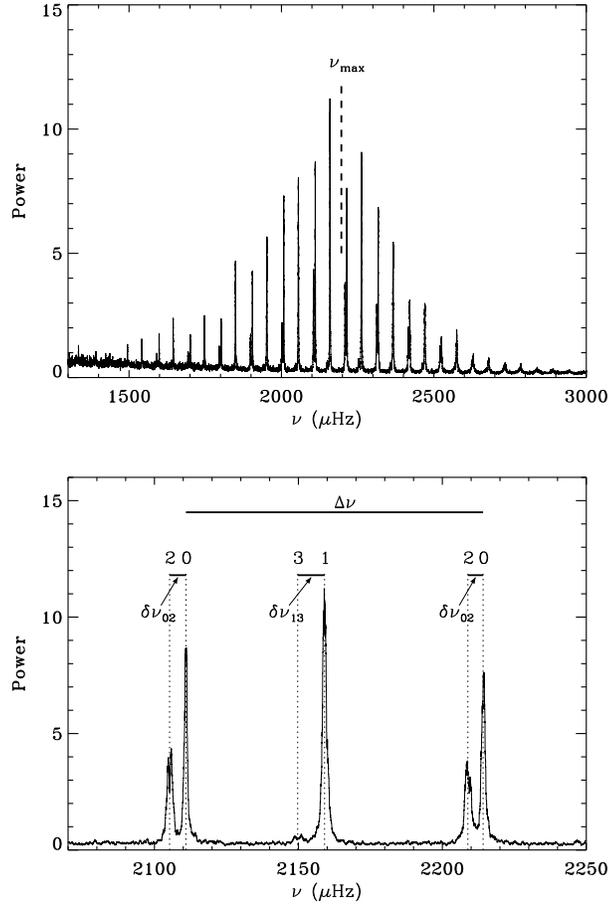}
\caption{Power spectrum of the solar-like pulsator 16 Cygni A,
from 35 months of {\it Kepler} data, smoothed with a $0.6 \muHz$ boxcar filter.
The lower panel shows a short segment;
the degrees of the modes and the large and small
frequency separations are indicated.
The inferred frequency $\nu_{\rm max}$ in the top panel, marked by a vertical
dashed line, and the fitted frequencies, marked by vertical dotted lines
in the lower panel, were obtained by Lund et al. (in preparation).
}
\label{fig:16cyga}
\end{center}
\end{figure}

The departure from the simple relation (\ref{eq:pasymp}) contains important
diagnostic information.
This is characterized by the {\it small frequency separation}
\begin{equation}
\delta \nu_{nl} = \nu_{nl} - \nu_{n-1\,l+2}
\simeq -(4 l + 6) {\Delta \nu \over 4 \pi^2 \nu_{nl}}
\int_0^R {\dd c \over \dd r} {\dd r \over r} \; ,
\label{eq:smallsep}
\end{equation}
where the last expression is valid only for main-sequence stars.
Here the integral is weighted towards the centre and hence is sensitive to
the sound-speed structure in the core.
For an approximately ideal gas, $c^2 \propto T/\mu$, where
$T$ is temperature and $\mu$ is the mean molecular weight; 
therefore $\delta \nu_{nl}$ is sensitive to the composition of the core,
and hence
to the amount of hydrogen that has been converted to helium by nuclear fusion,
determined by the age of the star.
As noted by e.g. \citet{Christ1988} this provides a simply way to determine
stellar ages, if other properties of the star are known.

The properties of the acoustic-mode spectrum in main-sequence stars are
illustrated in Fig.\ \ref{fig:16cyga} showing the power spectrum 
of the star 16 Cygni A as observed by {\it Kepler}.
The lower panel identifies the large and small frequency separations,
based on frequency fits to the power which in this case 
allows detection of modes of degree up to 3.

The second restoring force is gravity, acting through buoyancy on
density differences across horizontal surfaces;
consequently, this only operates for $l > 0$.
The resulting modes are standing internal gravity waves, or g modes.
They are characterized by the so-called buoyancy, or Brunt-V\"ais\"al\"a,
frequency $N$, given by
\begin{equation}
N^2 = g \left({1 \over \Gamma_1} {\dd \ln p \over \dd r} 
- {\dd \ln \rho \over \dd r} \right) \; ,
\label{eq:buoy}
\end{equation}
where $g$ is the local gravitational acceleration, $p$ is pressure,
$\rho$ is density and $\Gamma_1$ is the adiabatic compressibility.
In convection zones $N^2$ is negative, and hence the gravity waves are
evanescent.
In main-sequence stars the g modes have relatively low frequency, and
their detection in the Sun has been hotly debated for decades
\citep{Garcia2007, Appour2010}.
However, in evolved stars the gravitational acceleration, and hence $N^2$,
gets very high in the compact core of the stars, and hence g modes may
have high frequency, in the range of the solar-like p modes.
This gives rise to the very interesting phenomena of {\it mixed modes}
to which we return below.

As for p modes, the relevant g modes are often of high radial order, 
making their asymptotic behaviour of great diagnostic value.
This is most simply expressed in terms of the oscillation period
$\Pi = 1/\nu$ which approximately satisfies
\begin{equation}
\Pi_{nl} = \Delta \Pi_l (n + \epsilon_{\rm g}) \; ,
\label{eq:gasymp}
\end{equation}
where $\epsilon_{\rm g}$ is a phase that may depend on the degree.
Here the period spacing is
%$\Pi_l = [l(l+1)]^{-1/2} \Delta \Pi_0 $, where
\begin{equation}
\Delta \Pi_l = {2 \pi^2 \over  [l(l+1)]^{1/2}}
\left( \int N {\dd r \over r} \right)^{-1} \; .
\label{eq:perspac}
\end{equation}

For spherically symmetric stars the frequencies are independent of
the azimuthal order $m$.
This degeneracy is broken by departures from spherical symmetry, of which
by far the most important is rotation.
Rotation gives rise to a splitting which, for slow rotation, can
be written as
\begin{equation}
\nu_{nlm} = \nu_{nl0} + m \delta_{\rm rot} \nu_{nlm}
\label{eq:rotsplit}
\end{equation}
where $\delta_{\rm rot} \nu_{nlm}$ reflects an average of the internal
rotation rate, weighted by the properties of the oscillations.
In the solar case, this has allowed a detailed determination of 
solar internal rotation \citep[for a review, see][]{Howe2009}.
In the stellar case less information is obviously available,
but, as discussed below, some very interesting results have been obtained.

Solar-like oscillations are intrinsically damped but gain their energy
from the acoustic noise generated by the near-surface convection.
The result are peaks in the power spectrum with an amplitude that
is determined by the balance between the energy input and the damping and
a width that, for sufficiently long observations, is determined by the
intrinsic damping rate \citep[e.g.,][]{Christ1989}.
Early estimates of the energy input and the resulting amplitudes in 
the solar case were
made by \citet{Goldre1977}, while \citet{Christ1983} made a first estimate
of the amplitudes of solar-like oscillations across the relevant part
of the Hertzsprung-Russell diagram.
The damping is dominated by the effects of convection, involving the
perturbations to both the convective heat flux and the turbulent pressure
\citep[e.g.,][]{Balmfo1992}.
The treatment of these effects is highly uncertain, although various 
formulations of time-dependent convection have been established
\citep[see][for a review]{Houdek2015}.
With appropriate choice of parameters a reasonable fit can be obtained
to the observed solar line widths \citep{Chapli2005, Houdek2006}.

The combined result of the excitation and damping is a characteristic
distribution of power with frequency \citep{Goldre1994},
as shown in Fig.\ \ref{fig:16cyga} for the observations of 16 Cyg A.
This is characterized by the frequency $\nu_{\rm max}$ at maximum power.
There is substantial empirical evidence that $\nu_{\rm max}$ scales
as the acoustic cut-off frequency in the stellar atmosphere
\citep[e.g.,][]{Brown1991, Stello2008},
leading to $\nu_{\rm max} \propto M R^{-2} T_{\rm eff}^{-1/2}$, where
$T_{\rm eff}$ is the effective temperature.
The physical reason for this scaling has not been definitely established,
although \citet{Belkac2011} pointed out some likely relevant factors.

\section{Asteroseismic determination of stellar properties}
\label{sec:astero}

The space-based asteroseismic observations from CoroT and {\it Kepler}
have set the scene for
extensive investigations of stellar properties, ranging from ensemble studies
of large numbers of stars to detailed studies of individual targets.
These missions were, in part, motivated by the study of extra-solar
planetary systems (exoplanets) through the transit technique, with the
common requirement with asteroseismology 
of very high photometric precision over long periods of time.
We are still only at the beginning of exploring the potential of these
asteroseismic data,
and here I can just give a brief indication of the results obtained.
A recent review of asteroseismology based on solar-like oscillations
was provided by \citet{Chapli2013a}.

The most basic observed properties of solar-like oscillations are the
frequency $\nu_{\rm max}$ at maximum power and the large frequency
separation $\Delta \nu$.
These can be determined even from data with a low signal-to-noise ratio.
Assuming that $T_{\rm eff}$ is determined independently,
the scaling relations with acoustic cut-off frequency and stellar mean
density then provide two equations which can be solved for the mass and
radius \citep{Kallin2010}.
Even this simple analysis provides stellar quantities that are otherwise
very difficult to determine.
It can be refined by including constraints based on stellar model grids,
including also information about the stellar composition
\citep{Gai2011}.
A detailed test of these techniques was carried out by \citet{Silva2012}.
Alternatively, given the somewhat shaky foundations of the scaling for
$\nu_{\rm max}$, fits to grids of models can be carried out just based
on $\Delta \nu$ and $T_{\rm eff}$ \citep{Lundkv2014}.

These techniques provide simple methods to determine the basic properties
of large numbers of stars. 
They have been applied extensively to the CoRoT and {\it Kepler} observations,
and only a few examples can be given here.
\citet{Chapli2011, Chapli2014} analysed {\it Kepler} observations of
substantial samples of main-sequence stars,
in the early paper also comparing with the predicted distributions,
from Galactic modelling, in mass and radii of stars in the solar neighbourhood.
Data for huge numbers of red giants have been obtained by CoRoT
and {\it Kepler}, allowing detailed characterization of the population of
these stars \citep[e.g.,][]{Hekker2011}.
Analysis of stars in open clusters is particularly interesting.
Thus, based on {\it Kepler} data, \citet{Miglio2012} estimated the red-giant
mass loss in two open clusters from determination of stellar masses in
different evolutionary stages.

A very important application of basic asteroseismology of red giants is in
{\it Galactic archaeology}, relating stellar properties to the location
of the stars in the Galaxy \citep{Miglio2009}.
For red giants there is a close relation between stellar mass and age, 
and hence just the simple asteroseismic analysis provides a measure of
stellar age \citep{Miglio2013}.
When combined with large-scale spectroscopic investigations this provides
the basis for a detailed investigation of the chemical and dynamical
evolution of the Galaxy \citep[e.g.,][]{Casagr2016}.

When individual frequencies have been determined much more detailed and
accurate investigations of stellar overall and internal properties are
possible.
A difficulty in such analyses is the uncertain treatment of the near-surface
layers in the star and their effects on the oscillation frequencies, 
giving rise to what is known as the near-surface error in the computed 
frequencies.
In the solar case this can be isolated in the analysis owing to the
availability of observations over a large range of degrees.
Various techniques have been developed to correct for the effect in
distant stars, based on an assumed similarity with the solar correction
\citep[e.g.,][]{Kjelds2008, Christ2012} or with a somewhat stronger
physical basis \citep{Ball2014}.
Alternatively, model fits can be based on suitable ratios between
small and large frequency separation which are largely insensitive to
the near-surface effects \citep{Roxbur2003, Oti2005}.

An early analysis of {\it Kepler} data was carried out by \citet{Metcal2010},
for a star in the subgiant phase where hydrogen has been exhausted in the
core.
The resulting compact helium core increased the frequencies of gravity
waves in the deep interior, giving rise to modes of mixed p- and g-mode 
nature. 
The frequencies of such modes are very sensitive to the internal properties 
of the star, including its age, and hence the fit to the observed frequencies
in principle may result in very precise determinations of stellar properties.
However, several solutions were in fact found, each tightly constrained by
the data. 
This is an example of the importance of including additional information
about the star, supplementing the asteroseismic data,
as constraints on the stellar properties.
A detailed analysis of a sample of stars observed by {\it Kepler} 
was carried out by \citet{Mathur2012}
who also obtained some information about the dependence of the surface
correction on stellar properties.
Extensive modelling of two stars observed by {\it Kepler} was carried out by
\citet{Silva2013}, using several different modelling and fitting techniques
to test the range of systematic uncertainties involved in such fits.
Interestingly, one of the stars, with a mass of around $1.25 \Msun$,
had clear evidence for a convective core, with some additional mixing
outside the unstable region.
This is a first indication of the potential for using asteroseismology of
solar-like stars to study the physics of stellar interiors.
\citet{Metcal2015} fitted the full set of {\it Kepler} data for the two
components of the binary star 16 Cygni (see also Fig.\ \ref{fig:16cyga}).
As an encouraging test of consistency the independent analysis of the two
components yielded the same age, around 7\,Gyr, within the errors of
0.25\,Gyr, in accordance with the assumption of contemporaneous formation of
the pair.

Asteroseismology is playing an important role in the 
determination of properties of exoplanet host stars,
benefitting from the fact that the same photometric
observations can be used both for the characterization of the exoplanets and
for asteroseismology.
To determine the properties of an exoplanet we need the radius and mass
of the host star which can be determined much more accurately with 
asteroseismology than with `classical' astrophysical techniques.
Furthermore, asteroseismology allows determination of the age of the host
star and hence the planetary system.
In this way \citet{Batalh2011} found that {\it Kepler}'s first rocky
exoplanet, with a radius of $1.4 {\rm R_E}$, orbited a star with an age
of around 10\,Gyr, twice the age of the Sun.
Analysis of further data for this system by \citet{Fogtma2014}
yielded a value of the age of $10.4 \pm 1.4$\,Gyr and, remarkably, 
allowed a determination of the
radius of the planet with a precision of 125\,km.
A similar age was obtained by \citet{Campan2015} for a system containing
5 planets with sizes at or smaller than that of the Earth.
These striking results demonstrate that planet formation took place
already in the early phases of the history of the Galaxy.
I also note that a detailed analysis of a CoRoT exoplanet host was carried out
by \citet{Lebret2014}, who investigated the extent to which different 
combinations of seismic and non-seismic data could constrain the properties
of the star.

\begin{figure}[!]
\begin{center}
\includegraphics[width=8.7cm]{\fig/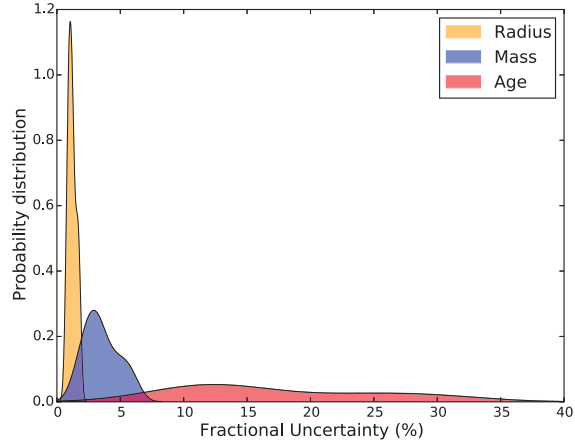}
\caption{Distributions of fractional uncertainties in asteroseismic
determinations of radius, mass and age for 33 {\it Kepler} (potential)
exoplanet hosts. Adapted from \citet{Silva2015}.
}
\label{fig:kages}
\end{center}
\end{figure}

\citet{Silva2015} carried out a detailed analysis of the 33 {\it Kepler} 
confirmed or potential exoplanet host stars for which extensive asteroseismic 
data are available. 
Taking into account also systematic effects of the use of different
modelling or fitting techniques, they were able to determine the radii
and masses with median uncertainties of 1.2 and 3.3 per cent, respectively,
whereas the ages were determined with a median uncertainty of 14 per cent.
The distributions of uncertainties are shown in Fig.\ \ref{fig:kages}.
I note that the age is determined predominantly from the decrease in the
central hydrogen abundance;
thus the fractional uncertainty in age, as illustrated, is unavoidably higher for
unevolved stars.

As mentioned in Section \ref{sec:properties} (see Eq.\ \ref{eq:rotsplit})
rotation causes
a splitting of the frequencies according to the azimuthal order $m$ which
in fact has been observed in a number of cases in the {\it Kepler} data.
The resulting information about the stellar rotation rate is obviously
of substantial interest.
However, in the exoplanet context an even more interesting aspect is
information about the orientation of the rotation axis.
For stochastically excited modes it is not unreasonable to assume that the
average amplitude, for given $n$ and $l$, is independent of $m$.
However, the {\it observed} amplitude depends on the inclination of
the rotation axis with respect to the line of sight \citep{Gizon2004}.
In the limiting case of a rotation axis in the plane of the sky 
only modes with even $l - m$ are observed, while if the rotation axis 
points towards the observer only modes with $m = 0$ are seen.
For exoplanets detected with the transit technique one would naively
expect the rotation axis of the host star to be in the plane of the
sky: the planets are assumed to form from a disk left over from the
formation of an initially rapidly rotating star and hence lying in the
star's equatorial plane; thus the rotation axis would be approximately
orthogonal to the plane of the planetary orbits, as is indeed the case
for the solar system.
Such systems have indeed been found \citep[e.g.,][]{Chapli2013b}.
However, in other cases there is a large misalignment between the rotation
axis and the axis of the planetary orbits 
\citep[e.g.,][]{Huber2013, Lund2014}.
Understanding the origin of this behaviour is an important part of the
study of the formation and evolution of planetary systems.

\section{Asteroseismology of red giants}
\label{sec:rg}

As a background to the discussion of the asteroseismology of red giants
it is useful to give a brief overview of red-giant evolution;
for a detailed review, see \citet{Salari2002}.
This phase of stellar evolution follows after the end of
central hydrogen burning.
The star continues to obtain its energy from hydrogen fusion, but now in
a shell around the gradually growing helium core.
The core contracts while the outer layers expand and cool, establishing
a deep outer convection zone.
When the star reaches the Hayashi track the continuing expansion
takes place at nearly constant effective temperature, leading to a drastic
increase in the luminosity (see also Fig.\ \ref{fig:pulshr}),
which in the solar case will reach as high
as one thousand times the present luminosity, along the red-giant branch.
At this point the temperature in the helium core has reached a level,
around 100 million degrees, where helium fusion to carbon and oxygen sets in.
The core expands and the outer layers contract, until the star settles down
to a phase of quiescent helium burning, a substantial fraction of the
energy still coming from the hydrogen shell burning.
After the end of central helium burning the outer layers again expand 
greatly in the asymptotic giant phase, after which the star sheds its 
envelope and is left with the central very compact carbon-oxygen core,
a white dwarf.

Given the deep outer convection zone it was expected \citep{Christ1983}
that red giants would show solar-like oscillations.
The first detection was made by \citet{Frands2002}, followed by a few
other ground-based studies which, however, were hampered by the very long 
observation periods required to resolve the low frequencies resulting from
the low mean density of the stars.
However, as already mentioned, a major break-through in the study of
solar-like oscillations in red giants came with the space-based observations 
from CoRoT and {\it Kepler} which have shown oscillations
in tens of thousands of stars.
The oscillations can be followed to the most luminous stars observed
by {\it Kepler}, with a power envelope similar to what is observed
on the main sequence (cf.\ Fig.\ \ref{fig:16cyga}) but with a dominant frequency
less than $1 \muHz$, corresponding to a period of more than 10 days
\citep{Stello2014}.
Indeed, the oscillations observed by {\it Kepler} merge with the even
slower oscillations seen in highly evolved giants with ground-based 
surveys \citep{Mosser2013}, and there is evidence that semi-regular
variables, with periods of many months and typically observed by
amateur astronomers, show solar-like oscillations \citep{Christ2001}.

From the first observations there were indications that the red-giant
modes had a very short damping time, leading to broad peaks in
the power spectrum and hence limited frequency precision, even though
an early theoretical estimate indicated life times several times the
typical solar values \citep{Houdek2002}.
Also, apparently only radial modes were observed.
This would further limit the diagnostic value of the observations.
A first indication of nonradial oscillations in a red giant
was obtained by \citet{Hekker2006} from observations of 
line-profile variations.
A definite proof that red giants showed the full range of solar-like
oscillations was obtained by \citet{Deridder2009} in an early analysis
of CoRoT data which also demonstrated the similarity of the power envelope
over a broad range of stellar luminosities and hence frequencies of
maximum power.

\begin{figure}[!]
\begin{center}
\includegraphics[width=8.7cm]{\fig/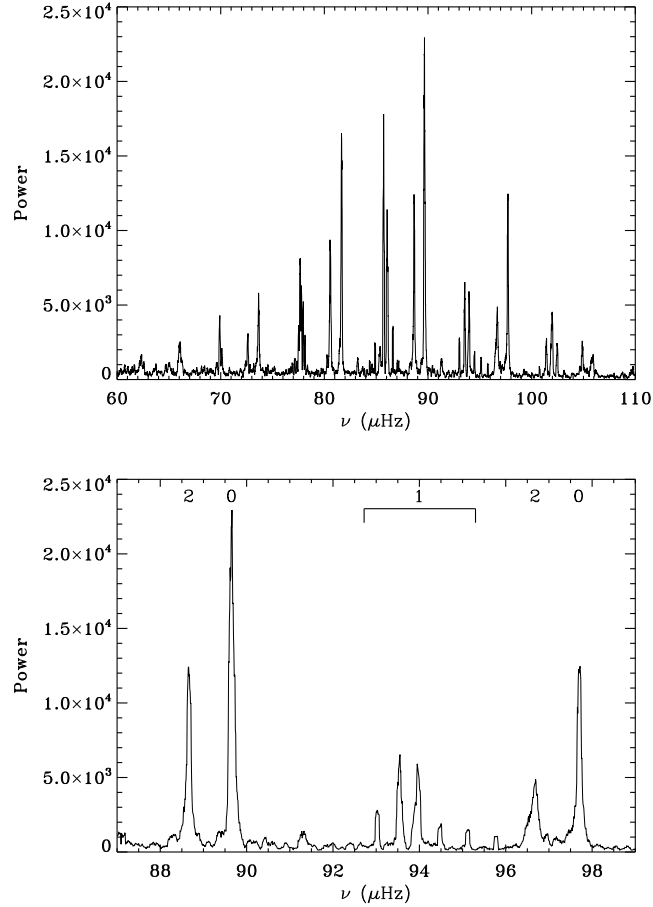}
\caption{Power spectrum from 49 months of {\it Kepler} observations
of the red giant KIC\,6779699, smoothed with a $0.09 \muHz$ boxcar filter.
In the small segment of the spectrum in the lower panel the degrees of the modes
are indicated, the horizontal line marking the $l = 1$ mixed modes.
See \citet{Beddin2011}.
}
\label{fig:powrg}
\end{center}
\end{figure}

Although the detection of non-radial modes in solar-like oscillations of
red giants was an important step,
the full, huge diagnostic potential of these observations became
apparent with the identification of mixed modes in a red giant by
\citet{Beck2011}, in {\it Kepler} observations.
This was followed by studies of ensembles of red giants by
\citet{Beddin2011} from {\it Kepler}, and \citet{Mosser2011} from CoRoT,
observations.
An example of an observed power spectrum is shown in Fig.\ \ref{fig:powrg}.
This is superficially similar (albeit at lower frequencies) to the
main-sequence power spectrum in Fig.\ \ref{fig:16cyga}, with pairs of
peaks of degree $l = 0$ and 2;
but instead of a single intermediate $l = 1$ peak there is now a group of peaks;
these are modes of mixed p- and g-mode behaviour.

\begin{figure}[!]
\begin{center}
\includegraphics[width=8.7cm]{\fig/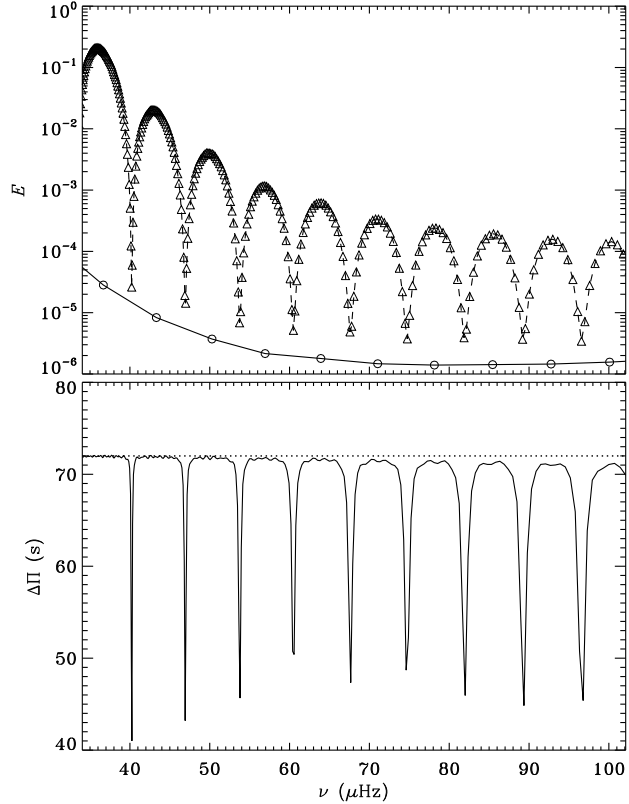}
\caption{Properties of oscillations in a red-giant model,
of mass $1 \Msun$ and radius $7 \Rsun$.
The upper panel shows the normalized mode inertia (cf.\ Eq. \ref{eq:inertia}) 
for modes of degree $l = 0$ (circles and solid line) and
1 (triangles and dashed line).
The lower panel shows the computed period spacings for $l = 1$,
the dotted horizontal line
marking the asymptotic value (cf.\ Eq.\ \ref{eq:perspac}).
}
\label{fig:perspac}
\end{center}
\end{figure}

To understand this behaviour it is instructive to 
consider the properties of modes in a stellar model,
characterized in terms of the normalized mode inertia
\begin{equation}
E = {\int_V \rho |\bolddelr|^2 \dd V \over M |\bolddelr|_{\rm phot}^2} \; ,
\label{eq:inertia}
\end{equation}
where the integral is over the volume of the star,
$\bolddelr$ is the displacement vector and $|\bolddelr|_{\rm phot}$ 
is its magnitude at the photosphere.
With this normalization $E$ is relatively small for modes trapped in
the outer parts of the star, whereas $E$ can be very large for modes 
trapped in the deep interior.
$E$ is plotted in Fig.\ \ref{fig:perspac}
as a function of frequency for modes of degree $l = 0$ and 1
in a red-giant model.
The radial modes are purely acoustic and have a small inertia that generally
decreases with increasing frequency.
The $l = 1$ modes are generally predominantly g modes, trapped in the deep
interior below the convective envelope, and hence have large inertia.
However, there are acoustic resonances 
where the inertia decreases to values not much
higher than the radial-mode inertia at the corresponding frequency.
Here the modes have their largest amplitude in the envelope where they
have an acoustic character.
The location of these resonances, and the radial-mode frequencies,
approximately satisfy the asymptotic relation in Eq. (\ref{eq:pasymp}).
Given that the processes exciting and damping the modes predominantly 
take place in the near-surface layers where the convective velocities
are large, it is intuitively clear that modes with low inertia are easier
to excite and hence are expected to be more visible in the power spectra
of the observations \citep[see][]{Dupret2009, Grosje2014}.
This is the origin of group of $l = 1$ peaks in Fig.\ \ref{fig:powrg};
these are modes with inertia somewhat higher than the radial-mode inertia,
but still excited to observable amplitudes.

Given the asymptotic behaviour of g modes (Eq.\ \ref{eq:gasymp})
the properties of the mixed modes are most naturally analysed in terms of
period spacings which, as shown in the lower panel of Fig.\ \ref{fig:perspac},
are also affected by the acoustic resonances.
For the modes of predominantly g-mode character the
spacing $\Delta \Pi = \Pi_{nl} - \Pi_{n-1\,l}$ is close to the
asymptotic value (cf.\ Eq.\ \ref{eq:perspac}) shown
by the horizontal line.
However, at the acoustic resonances the period spacing takes on a 
characteristic `V'-shape as a function of frequency,
a behaviour that led \citet{Beck2011} to the first
identification of mixed modes in red-giant observations.

From the observed frequencies of the mixed modes one can determine the 
period spacings around the acoustic resonances and, most reliably from
a fit to the detailed asymptotic behaviour of the frequencies
\citep{Mosser2012a},
determine the asymptotic period spacing $\Delta \Pi_l$
(Eq.\ \ref{eq:perspac}).
It was shown by \citet{Beddin2011} and \citet{Mosser2011} that the
period spacing provides a clear separation between otherwise very similar
stars ascending the red-giant branch with just shell hydrogen fusion and
stars in the core helium-fusion phase:
the period spacing was substantially smaller in the former case than in the
latter.
This can be understood from Eq. (\ref{eq:perspac}), according to which
the asymptotic period spacing is determined by an integral over
the buoyancy frequency (Eq.\ \ref{eq:buoy}).
When the star moves to the core helium-burning phase the core expands,
and this decreases the local gravitational acceleration and hence
the buoyancy frequency.
A further reduction of the integral results from the convective core caused by
helium fusion, since the integration in Eq. (\ref{eq:perspac}) excludes the
convective core. 
Both effects decrease the magnitude of the integral 
and hence increase the asymptotic period spacing.
Combining the period spacing with the large frequency separation $\Delta \nu$,
which varies strongly with stellar radius,
allows detailed diagnostics of stellar evolution, as discussed by
\cite{Mosser2014}.

Further information about stellar interior structure,
such as the properties of the
convective core in helium-burning stars, may in principle be obtained
from detailed fitting of the individual oscillation frequencies.
Although such fits have been attempted in a few cases
\citep[e.g.,][]{Dimauro2011, Jiang2011} much work is still required to
explore these possibilities.

\begin{figure}[!]
\begin{center}
%\vspace{-7.5cm}
\hspace{1cm}\includegraphics[width=9.5cm]{\fig/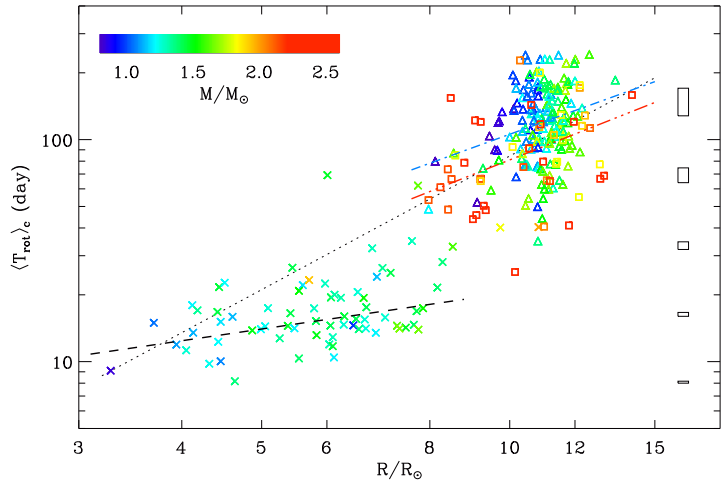}
%\vspace{8.0cm}
\caption{Asteroseismically inferred core rotation periods in
red giants (crosses) and core helium-burning stars (triangles and squares),
plotted against stellar radius in solar units.
The colour code indicates stellar mass.
The right-hand boxes show typical errors, depending on the period.
From \citet{Mosser2012b}.
}
\label{fig:rotrg}
\end{center}
\end{figure}

From a determination of the rotational splitting (cf.\ Eq.\ \ref{eq:rotsplit})
in {\it Kepler} observations
of a red giant \citet{Beck2012} concluded that the core of the star rotated
faster than the surface by around a factor 10.
This was based on determining the splitting for mixed modes, including
modes with a substantial g-mode component where the splitting was dominated
by the core.
Fast core rotation was also found through asteroseismic inversion 
in less evolved stars, in the sub-giant phase
and near the base of the red-giant branch,
by \citet{Deheuv2012, Deheuv2014} and \cite{Dimauro2016}.
As shown in Fig.\ \ref{fig:rotrg}
\citet{Mosser2012b} determined the core rotation of a large number
of stars on the red-giant branch and in the core helium-burning phase.
Combined with the nearly uniform rotation inferred in the solar interior
\citep[cf.][]{Howe2009} and a recent asteroseismic determination of overall
rotation in old field stars \citep{vanSad2016}
these results provide indications of the evolution of
stellar interior rotation with age, a process that may also have important
consequences for stellar structure evolution as a result of related 
instabilities and mixing processes.

In fact, the rapid core rotation in red giants should come as no surprise.
As discussed above the evolution on the red-giant branch involves a strong
contraction of the core.
If there were local conservation of angular momentum this would result a
spin-up of the core to far higher rotation rates on the red-giant
branch than in fact inferred from the asteroseismic determinations.
Thus some angular-momentum transport mechanism must be operating in
the stellar interior, leading to a reduction of the angular momentum 
and hence the rotation rate in the core.
The normally considered mechanisms for angular-momentum transport 
in stellar interiors are somewhat uncertain. 
However, it has been found that they are insufficient, by one to two orders
of magnitude, to account for the observed rotation in red-giant stars
\citep[e.g.,][]{Eggenb2012, Marque2013, Cantie2014}.
Thus additional transport mechanisms are required.
Internal gravity waves
\citep{Fuller2014} or mixed modes \citep{Belkac2015} may play an important role.
Even so, it is clear that we are still not close to understanding these
important aspects of stellar evolution.

An early mystery in the study of solar-like oscillations in red giants
was the suppression of the $l = 1$ modes in some stars with otherwise
apparently normal oscillation spectra \citep{Mosser2012c, Garcia2014}.
\citet{Stello2016} demonstrated that these stars had a mass, slightly
higher than the Sun,
such that they would have had convective cores on the main sequence.
On this basis \citet{Fuller2015} proposed that the $l = 1$ modes were
suppressed by scattering by a fossil magnetic field in the core of the star,
generated through dynamo action when the star was on the main sequence.
Although this model needs to be tested through more detailed calculations,
it represents yet another instance of the power of asteroseismology to probe
the evolution of these evolved stars.

\section{Future prospects}

%\note [K2, TESS, PLATO. \citet{Chapli2015}]

The CoRoT mission ended operations in December 2013 after 7 years
and the {\it Kepler} nominal mission ended in the spring of 2013 with
the breakdown of two of its four reaction wheels.
This, however, is far from the end of space asteroseismology.
Operations of {\it Kepler} are continuing in the K2 mission, where
successive fields along the Ecliptic are observed for three-months periods
\citep{Howell2014}.
With this orientation stable pointing of the satellite can be achieved with
just the remaining two reaction wheels.
Early results from this mode of operation are promising, both for
asteroseismology of stars near the main sequence \citep{Chapli2015} and
for red-giant observations as applied to Galactic archaeology
\citep{Stello2015}.
This will be followed by the TESS mission 
\citep[Transiting Exoplanet Survey Satellite,][]{Ricker2014} scheduled
for launch by NASA in 2017.
Over a two-year period TESS will make a survey of nearly the entire 
sky, to search for exoplanets and carry out asteroseismology. 
Most fields will be observed for around 28 days, but for two fields at the
ecliptic poles the observations will be continuous for a year each.
A major advantage of TESS compared with {\it Kepler} is the focus on relatively
nearby stars which greatly enhances the possibilities for supplementary
ground- and space-based observations. 
This advantage is shared by ESA's PLATO mission \citep{Rauer2014},
selected for launch in 2024.
PLATO will observe fields much larger than {\it Kepler}'s, in two
cases for two or three years, emphasising the characterization of
Earth-like planets in the habitable zone.
Asteroseismology will be possible for a large fraction of the exoplanet
candidates detected and, obviously, for a large number of other stars.

In parallel with the exciting prospects offered by these new missions,
the analysis of the CoRoT and nominal {\it Kepler} data has far from been
completed.
Indeed, efforts to go beyond the basic characteristics of the stars
are just starting.
An important example is the characterization of convective core overshoot
in main-sequence stars from analysis of solar-like oscillations
\citep{Deheuv2015a}.
The evolution of rotation with age will remain a key topic of research.
Interestingly, from {\it Kepler} data \citet{Benoma2015} 
found a general tendency to near-uniform rotation in main-sequence stars,
as has also been found in the Sun from helioseismology,
and in strong contrast to the rapid core rotation on the red-giant branch.
Moving to later evolutionary stages \citet{Deheuv2015b} found little 
radial variation of rotation in core helium-burning stars.
Thus we see the first signs of an overall characterization
of rotation in the different evolutionary stages.

Despite the success of space-based asteroseismology, ground-based 
observations should not be ignored.
In fact, solar observations have demonstrated that the intrinsic
stellar background `noise' from near-surface convection and activity
is a much more serious concern in photometric asteroseismic observations than
in radial-velocity observations.
This is the motivation for the creation of the Danish-led 
SONG (Stellar Observations Network Group) network of 1\,m telescopes
dedicated to asteroseismology and exoplanet studies \citep{Grunda2014}.
The first telescope in the network, the Hertzsprung SONG Telescope, is
in operation on Tenerife, and the second telescope is in commissioning in
Delingha in western China. 
Funding for further telescopes will be sought from Danish sources and
through international collaboration.

Thus, referring again to Eddington, there are excellent prospects to
obtain certain, or at least much improved, knowledge of that which is hidden 
behind the substantial barriers of the stellar surface.

%\begin{align}
%\label{perturb} 
%y &= y_0 + y^\prime \\
%\left| y^\prime \right| &\ll \left| y_0 \right|   
%\end{align}
%
%\begin{eqnarray}
%\label{wave_equation}
%\left(\derivp{^2}{t^2} - \vec L \right) \vec v_{\rm osc} + {\vec  {\cal C}_{\rm osc}} = \vec {\cal S}_t\,,
%\end{eqnarray}
%
%\begin{align}
%\label{Source}
%\mathcal{S}_t &= \mathcal{S}_{R} + \mathcal{S}_{E} + \mathcal{S}_\Omega +  \mathcal{S}_M + \mathcal{L}_t
%\end{align}

\section*{Acknowledgment}
I am grateful to the organizers for the invitation to participate 
in this very interesting conference and the opportunity to visit Azerbajan.
I thank Rasmus Handberg for help with producing Figs \ref{fig:16cyga} and
\ref{fig:powrg}, Victor Silva Aguirre for providing Fig.\ \ref{fig:kages}
and Beno\^{\i}t Mosser for providing Fig.\ \ref{fig:rotrg}.
Funding for the Stellar Astrophysics Centre is provided by 
The Danish National Research Foundation (Grant DNRF106). 
The research is supported by the ASTERISK project
(ASTERoseismic Investigations with SONG and Kepler) 
funded by the European Research Council (Grant agreement no.: 267864).

%\medskip
%\,

\end{document}